\documentclass{article}

\usepackage{arxiv}

\usepackage[utf8]{inputenc} 
\usepackage[T1]{fontenc}    
\usepackage{hyperref}       
\usepackage{url}            
\usepackage{booktabs}       
\usepackage{amsfonts}       
\usepackage{nicefrac}       
\usepackage{microtype}      

\usepackage{graphicx}

\usepackage{doi}
\usepackage{siunitx}
\usepackage{multirow}
\title{Phonon Interference at the Atomic Scale }

\author{
	 \href{0000-0001-7257-8125}{\includegraphics[scale=0.06]{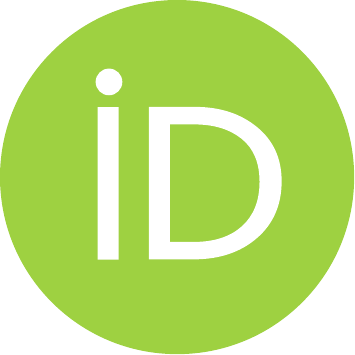}\hspace{1mm}Paul Desmarchelier} \\
	Univ Lyon, INSA Lyon, CNRS,CETHIL, UMR5008, 69621 Villeurbanne, France\\
	\texttt{paul.desmarchelier@insa-lyon.fr} \\
	\And
	Efstrátios Nikidis \\
	 Physics Department, Aristotle University of Thessaloniki, GR-54124 Thessaloniki, Greece\\
	\And
	Yoshiaki Nakamura\\
	OSAKA University, 560-8531 Osaka, Japan\\
	\And
	Anne Tanguy\\
	Univ Lyon, INSA Lyon, CNRS, LaMCoS, UMR5259, 69621 Villeurbanne, France\\
	ONERA, University Paris-Saclay,Chemin de la Hunière, BP 80100, 92123 Palaiseau, France\\
	\texttt{anne.tanguy@insa-lyon.fr} \\
		\And
\href{0000-0002-6933-2674}{\includegraphics[scale=0.06]{orcid.pdf}\hspace{1mm} Joseph Kioseoglou}\\
	Physics Department, Aristotle University of Thessaloniki, GR-54124 Thessaloniki, Greece\\

 \AND
\href{0000-0002-8521-7107}{\includegraphics[scale=0.06]{orcid.pdf}\hspace{1mm}Konstaninos Termentzidis} \\
Univ Lyon, CNRS, INSA Lyon,CETHIL, UMR5008\\
69621 Villeurbanne, France\\
\texttt{konstantinos.termentzidis@insa-lyon.fr} \\
}

\renewcommand{\shorttitle}{ Phonon Interference at the Atomic Scale}

\hypersetup{
pdftitle={Phonon Interference at the Atomic Scale},
pdfsubject={cond-mat, nanomaterials},
pdfauthor={Paul Desmarchelier},
pdfkeywords={Interference, Diffraction, phonon },
}

\begin{document}
\maketitle

\begin{abstract}
Phonons diffraction and interference patterns are observed at the atomic scale, using molecular dynamics simulations in systems containing crystalline silicon and nanometric obstacles as voids or amorphous-inclusions. The diffraction patterns caused by these nano-architectured systems of the same order as the phonon wavelengths are similar to the ones predicted by a simple Fresnel-Kirchhoff integral, with a few differences due to the nature of the obstacle and the anisotropy of crystalline silicon. These findings give evidence of the wave nature of phonons, can help to a better comprehension of the interaction of phonons with nanoobjects and at long term can be useful for intelligent thermal management and phonon frequency filtering at the nanoscale.

\end{abstract}

\keywords{Interference, Diffraction, phonon}

Diffraction and interference can be observed for any wave propagation and has been studied since the sixteen century for light \cite{Huygens1920}, the most famous example being the double slit experiment \cite{Young1802}. Diffraction is well described since the end of the nineteenth century through the Fresnel-Kirschoff equation \cite{Kirchhoff1883}.
It is very commonly used in optics, but has applications in acoustics and geophysics \cite{Deregowski2006}.
Indeed, this model is valid for any kind of propagating wave, including electromagnetic waves (photons) and lattice vibrations (phonons). More broadly, this can be applied to the quantum wave description of particles \cite{Marcella2002}. Hong–Ou–Mandel interference have even been observed for phonons \cite{Toyoda2015}.

Nevertheless, the first experiment showing  phonons interference or more largely interference of high frequency propagative waves in solids date from the second half of the last century. The work of Anderson and Sabisky is one of the first describing phonon interference~\cite{Anderson1970}. Since then, interferences between reflected and incident waves have been used to engineer phonon bandgaps in superlattices or phononic crystals \cite{Xie2014}, or to induce interference by creating two phonon pathways that will then interact with each other, decreasing the transmission of specific frequencies~\cite{Hu2020}.
More globally, the importance of the wavelike nature of phonons, is highlighted in recent reviews \cite{Simoncelli2019,Zhang2022}. Controlling interference patterns allows managing the spatial extent of energy transfer.

Phonon diffraction, when defects change the direction of propagation of phonons, can also affect the lattice heat conduction. It has been shown that dislocations can act as a diffraction grating \cite{Hanus2018}. The diffraction phenomena have also been evidenced thanks to the interference patterns induced by periodic transducers in pump-probe experiments, as evidenced by the influence on conduction \cite{Vu1995} or visualization of the pattern via angle resolved Brillouin scattering \cite{Dieleman1999}.
Finally, diffraction by a small aperture has been shown experimentally \cite{Zaitlin1975} and described theoretically \cite{Hanus2018}.


It is challenging to observe spatially resolved interference and diffraction of phonons in the THz range directly, usual methods consisting in using indirect effects, such as on the heat flux~\cite{Vu1995}. This limitation can be waved by using Molecular Dynamics simulations that give access to the different quantities at the atomic scale, allowing the direct visualization of high-frequency waves. 

\begin{figure}[h] 
	\centering
	\includegraphics[width=.5\linewidth]{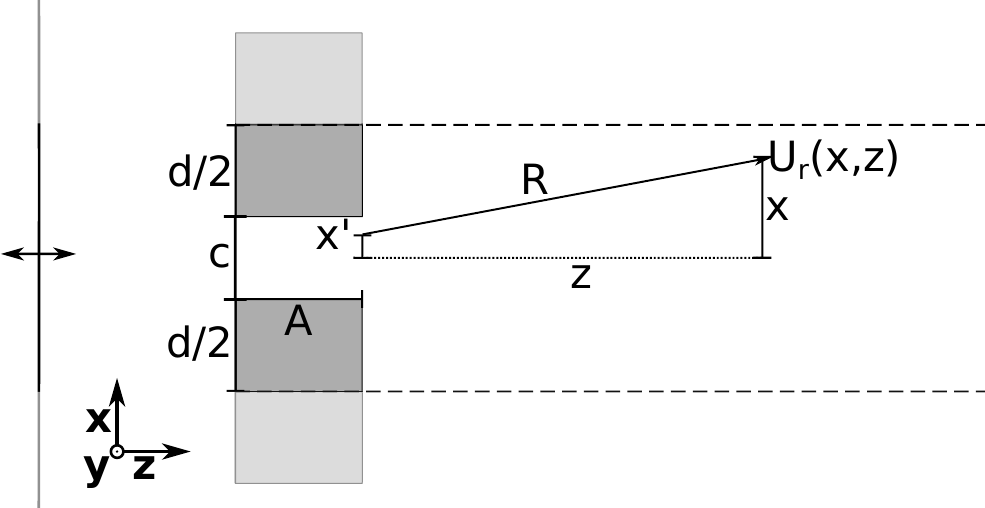}
	
	\caption{ \label{fig:Schemekir} Geometrical description used for the computation of the displacement field due via equation [\ref{eq:kirch}] } 
\end{figure}

In this letter, we show diffraction and interference of the THz range phonons induced by voids or by amorphous patches. For this, we study the propagation of a longitudinal wave in a single crystal of Si separated in two parts, with a small interconnecting channel. The Si block has an orientation $\langle 100 \rangle$ in the $z$ direction, from this block of thickness $A=$ 8 nm is removed to the exception of a small channel (see Fig. \ref{fig:Schemekir}). This channel will be our aperture to diffract the incoming wave. In other words, the diffraction grating is build using nanometric crystalline bridges over a void. As diffraction occurs when the size of the obstacle is of the order of the wavelength, the channel width is set to 3 \si{\nm}. For reference, the wavelength in c-Si at 6 \si{\THz} is 1.2 \si{\nm} (for longitudinal phonon in the $\langle 100 \rangle$ direction). This value is extracted from the dispersion relation for the potential used, using the dynamical structure factor (the exact method is described in a previous article \cite{Desmarchelier2021}).
A void barrier is used here, but any material having a high acoustic mismatch or a strong attenuation in the THz range can be used, such as amorphous Silicon (see \ref{app:a-Si}). 
Periodic boundary conditions are used in all the directions except the wave propagation direction. As a result, we obtain a transmission diffraction grating of infinite size, the slits (here channels) are infinitely long in $y$ and repeated infinitely in $x$. The  diffracted wave exiting the channel interacts with its image wave through the periodic boundary condition. To depict this phenomenon, we have chosen a rather large length after the channel exit (here 111 \si{\nm}). The dimensions of the simulation box are of 250 \si{\nm} in length, 47 \si{\nm} in width and 4 \si{\nm} in thickness. The total length of 250 \si{\nm} is chosen to let the wave propagate in both $-z$ and $+z$ direction to avoid the waves interacting with each other through reflection. The exact dimensions are adjusted to have an integer number of lattice units and avoid the creation of crystal defects. At the edge of the box, in the $z$ direction, the atoms are frozen to avoid interaction between the two directions of propagation ($-z$ and $+z$). The interatomic potential used is the one designed by Vink et al. \cite{Vink2001} and LAMMPS is used to run the simulation \cite{Plimpton1997}.

\begin{figure}[h] 
	\centering
	\includegraphics[width=\linewidth]{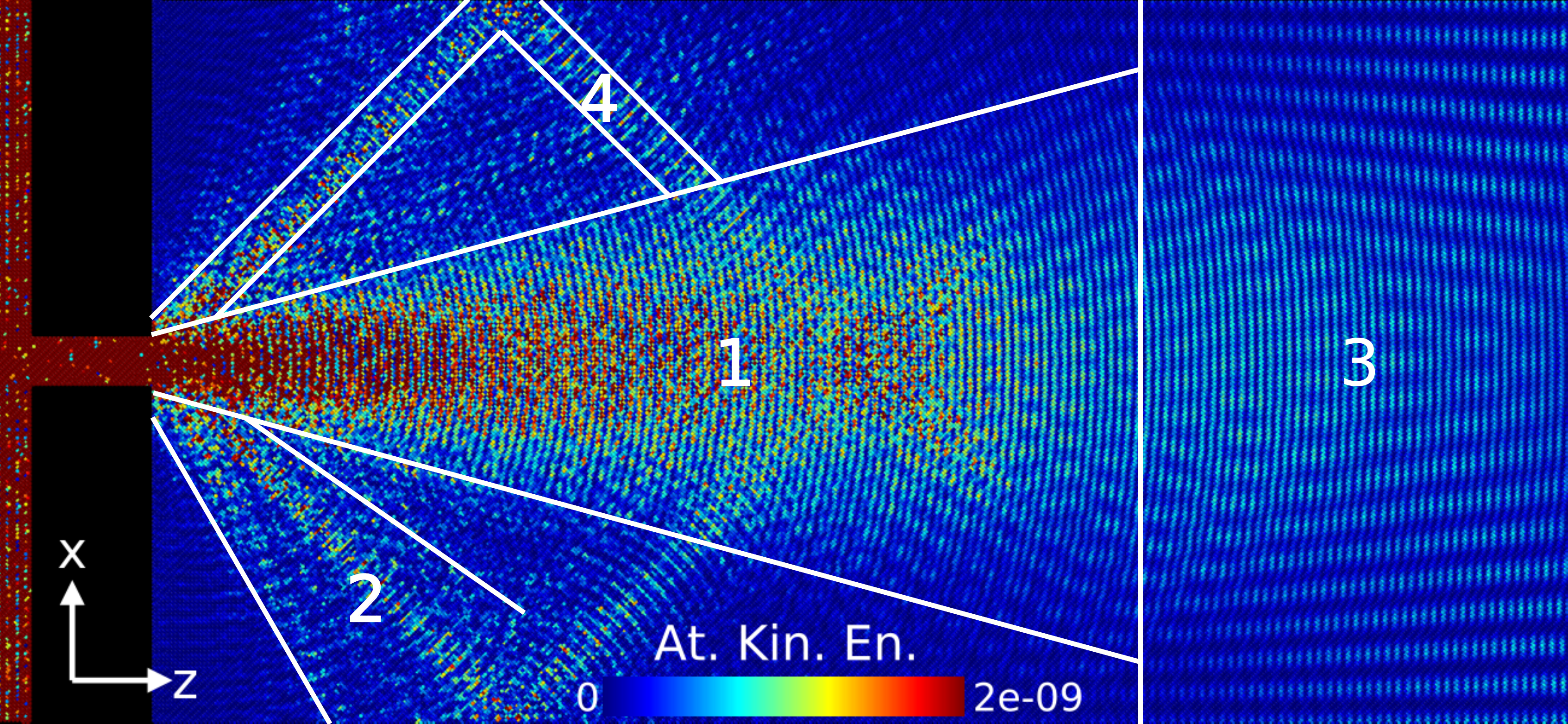}
	
	\caption{ \label{fig:DiffVisu} Visualization of diffraction of a 6 THz longitudinal plane wave by an aperture of 3 nm in width. The color scale represents the kinetic energy per atom in eV. Here only  a small part of the  section before and the section after the channel of the simulation box is depicted.} 
\end{figure}

The wave used to visualize the diffraction interference is created by exciting a 4 \si{\angstrom} thick slice with a sinusoidal continuous force excitation in the $z$ direction, represented in Fig. \ref{fig:Schemekir} by the vertical black line on the left with the arrows giving the polarization. For this work, the frequency chosen is 6 THz that has a wavelength of 1.2 \si{\nm}, corresponding to the order of the aperture size. The amplitude of the excitation is very low and results in local displacement of \SI{1e-4}{\angstrom} at most, such a low excitation amplitude is possible because the system is at mechanical equilibrium a the begging of the simulation (T= 0K). This low excitation amplitude is necessary as the system is continuously excited and thus energy is continuously accumulated in the system. An atomic visualization of the system is displayed in figure \ref{fig:DiffVisu}. In this figure different features appear, the most obvious one, labeled 1 is the slow broadening of the waves front at the exit of the channel due to diffraction. The second feature, labeled 2, are the side lobes. These side lobes are centered around \ang{35} and are typical for diffraction by an aperture and will be discussed in the following paragraphs. In the last fifth of the picture, labeled 3, a pattern composed of slightly slanted lines appear, these lines build the beginning of the well known far field interference pattern. Finally, in the zone labeled 4, lines forming an angle of \ang{45} with the $z$ axis appear, these lines disappear when the void is filled with a-Si (see figure \ref{app:a-Si}). 
To facilitate the simulation, the kinetic energy is used to visualize the shape of the waves rather than the displacement. It is directly correlated to the displacement (proportional to the square of the derivative).


To confirm that the patterns observed here are caused by diffraction and interference of the incoming waves, the results of the MD simulations are compared with a continuous model of wave propagation. The model of diffraction used is the Fresnel-Kirschhoff equation~\cite{Peatross2015} assuming a superposition of emitted spherical waves:
\begin{equation}\label{eq:kirch}
U_r(x,z)=A\int_{-c/2}^{c/2} U_r(x',0)\frac{exp(ikR)}{R}\left[\frac{1+R/z}{2}\right] dx'
\end{equation}
with $dx'$ the integral over the aperture, $R=\sqrt{(x-x')^2+z^2}$, and $k=2\pi/\lambda$ ($\lambda$ being the wavelength). The different geometrical parameters are defined in Fig. \ref{fig:Schemekir}.  The origin of the $z$ axis is set at the aperture opening, the origin of the x axis is taken at the center of aperture. To compute the full displacement field, the integral is computed for each point in a grid of dimensions corresponding to the MD simulations, with a resolution of dz=1.6 \si{\angstrom} and dx=1.1 \si{\angstrom}. The integral is solved using a Clenshaw–Curtis method implemented in SciPy \cite{SciPy}. $U_r(x',0)$ is here defined as a constant which is equivalent to assume a uniform displacement at the exit of the channel. The interface effects in the channel are neglected. Another strong assumption of equation [\ref{eq:kirch}] is that wave-propagation in c-Si is isotropic.
Moreover, as this represents a sinusoidal displacement field, a quantity proportional to the kinetic energy can be obtained through the square of the displacement: 
\begin{equation}\label{eq:EkKir}
E_k(x,z)\propto	U^2_r(x,z)
\end{equation}
To obtain the interference pattern, a set of apertures must be considered. For this, the solution in terms of displacement field of the diffraction by a single aperture is simply shifted by c+d in the x direction and summed.
It is noteworthy that here, due to the intermediate size of the system, the small angle approximation does not hold, the neither the Frauenhofer nor the Fresnel approximation can be used \cite{Peatross2015}.

\begin{figure}[h] 
	\centering
	\includegraphics[width=.48\linewidth]{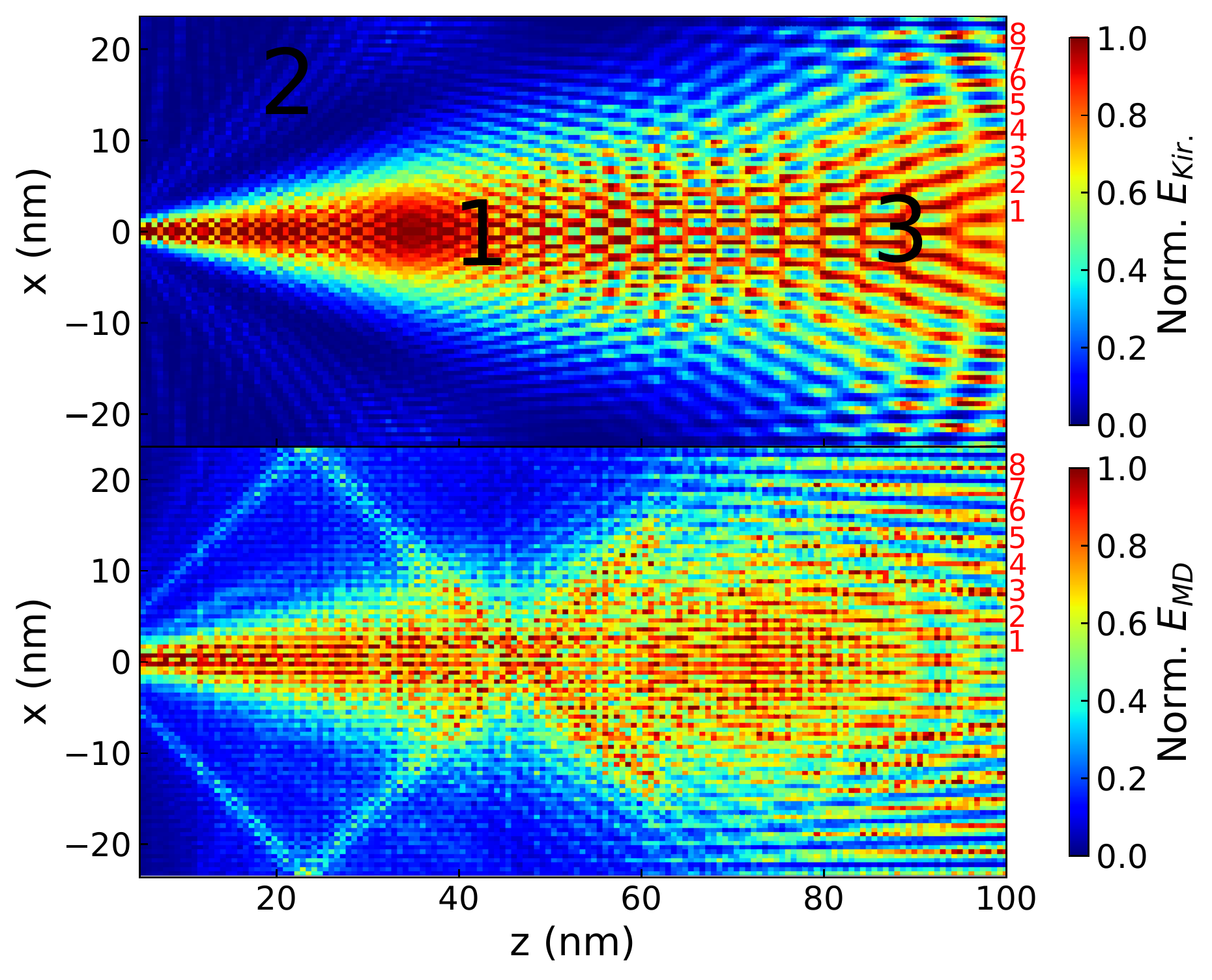}
	\includegraphics[width=.48\linewidth]{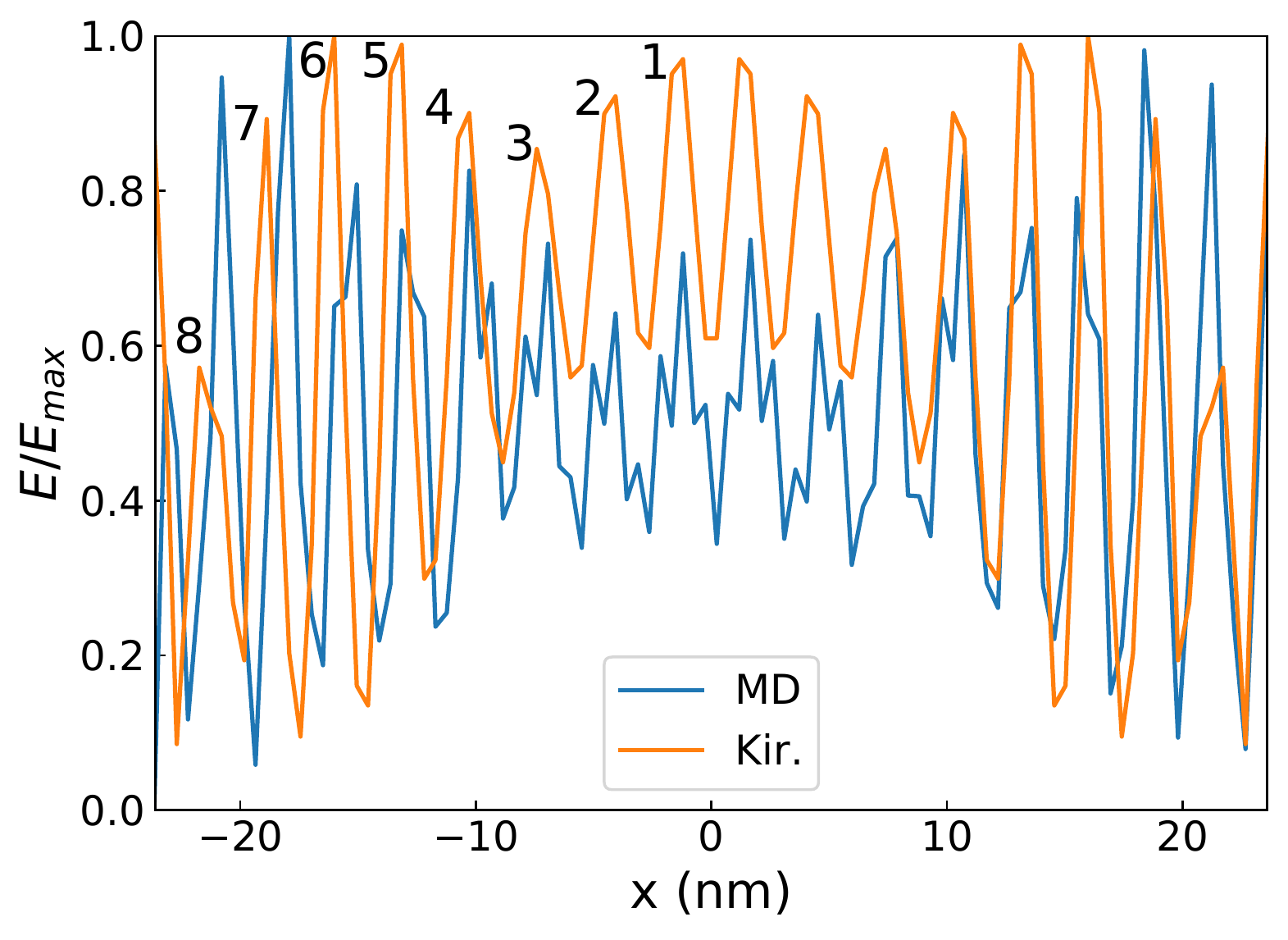}
	
	\caption{ \label{fig:Comp} Comparison of the results obtained with the Kirchhoff integral (top left) and the MD simulation (bottom left) as a heat map, and for a slice at z=100 nm (right). The energy value is normalized by the maximum value for each pixel column along z. } 
	
\end{figure}

The pattern obtained with this method for the studied geometry is displayed in Fig. \ref{fig:Comp} (top left). The aperture is centered around x= 0. To ease the comparison with the MD simulation, the results of the simulation are also reported in a panel below the results of the Fresnel-Kirchhoff equeations. For both panels, the color scale is normed, for each pixel column along z, by the highest value in the pixel column. For completeness, a figure without such a normalization process is displayed in the appendix \ref{app:Log}. The same zones identified in the atomistic representation (see Fig. \ref{fig:DiffVisu}), are present here. The zone 1, where the slow broadening of the central part due to the diffraction is visible. The zone 2, where a secondary lobe at around \ang{35} appears, and the zone 3, where an interference pattern in the form of slanted lines are visible. To compare directly with the MD results in the panel below, these lines are numbered: around the symmetry plane at x=0 there are 8 fringes on each side 100 nm away from the aperture.
For a more precise comparison, the energy distribution as a function of x for z = 100 nm is displayed on the right of Fig. \ref{fig:Comp}. The broad pattern of having more energy in the center and less energy at the edges due to the simple diffraction is visible for both curves. Again, one can verify that there is the same number of fringes in both cases. 
This similarity with the models is a sign that equation [\ref{eq:kirch}] reproduces well the patterned observed with the MD simulation.

Nevertheless, we observe a few differences between the patterns obtain through equation [\ref{eq:kirch}] and the one obtained with the MD simulation. The first obvious are the lines at \ang{45} appearing for MD, (labeled 4 in Fig. \ref{fig:DiffVisu}). As they disappear when the voids are filled with a-Si and are attenuated for shorter channels, they can be attributed to the effect of the free surface in the channel, most probably reflection (figures \ref{fig:Am} and \ref{fig:SmallerChannel}). 
When looking at the detail at z=100nm it appears that even-though the number of fringes between the two models matches, their positions are phase shifted. This might be linked to the anisotropy being neglected in the model. Indeed, crystalline Si is anisotropic \cite{Hopcroft2010}, this role of anisotropy can be verified in the appendix \ref{app:or} where we show that an orientation change affects slightly the observed pattern. This anisotropy is not included in equation \ref{eq:EkKir} that assumes that the medium is isotropic. The treatment of anisotropy for diffraction in the near field exceeds the scope of this work. Moreover, here the constant frequency surface does not deviate much from a sphere for silicon in the direction studied \cite{Thomsen1986,Wolfe1998}.    
It also appears clearly that the spatial pattern of the kinetic energy resulting from MD simulations is less sharp and contrasted than the one using the equation [\ref{eq:kirch}]. This can be explained by the continuous medium model that has been used for the implementation of the wave propagation which contrasts with the discrete atomistic nature of the materials and the MD simulations. 
Also, due to the normalization by a very low value of kinetic energy, the appearance of numerical noise in the background is possible.
Finally, the norm used here hides a discrepancy between the two models, the attenuation rate in the analytical model is much stronger than the one observed with molecular dynamics (see figure \ref{fig:CompLog}): the channel present in MD appears thus to focus the energy.



It is worth to keep in mind that the simulations that have been performed here are at very low temperature with the sinusoidal excitation as the only source energy, in order to minimize the thermal noise. This facilitates the visualization of the results, nevertheless since the length scale involved here are lower than the mean free path in Si, we expect that the phenomena happens also at a finite temperature. However, at higher temperature the kinetic energy used to visualize the interferences would be dominated by noise.
The pattern observed may also be discussed, as is usual for diffraction one might expect to observe pattern described by a cardinal sine function. However, in the present study we are not in the far field and deal with an infinite grating, those two factors may prevent such a solution to arise.

To conclude, we have shown using MD that diffraction and interference patterns appear for waves at frequencies important for thermal transport in silicon. These interference patterns can be reasonably predicted using the Fresnel-Kirchhoff equation but with discrepancies, that depend on the nature of the boundary obstacles (voids or amorphous inclusions, size of the channel, orientation of the crystal). Due to its simplicity, the model is robust and might be applied at larger or smaller scale for other frequencies, but it ignores the specificities of surface and anisotropy effects that are detailed in the letter. This study indicates that the wave nature of phonons matters even at high frequencies in crystals. Interestingly, amorphous patches in a crystalline Si sample are sufficient to induce diffraction and then interferences between phonons. The wave nature of high frequency phonons might be important for phonon focusing applications, where phononic crystal are studied using an approach that considers phonons as particle \cite{Anufriev2020}. Finally, phonon diffraction grating may help to select phonon of specific frequencies.
\section*{Acknowledgments}

	This work was granted access to the HPC resources of IDRIS under the allocation 2021-
	A0110911092, made by GENCI, and also granted access from the Greek Research and Technology Network (GRNET) in the National HPC facility -ARIS- under the project NOUS(pr012041).This work was supported in part by NSF Grant No. PHY-1748958 and the Gordon and Betty Moore Foundation Grant No. 2929.02. Authors want to thank fruitful discussions with Carsten \textsc{Henkel} and Gabriel \textsc{Dutier}.

\appendix

	\section{Comparison in Non-Normalized Values}
	\label{app:Log}
\begin{figure}[h] 
		\centering
		\includegraphics[width=.5\linewidth]{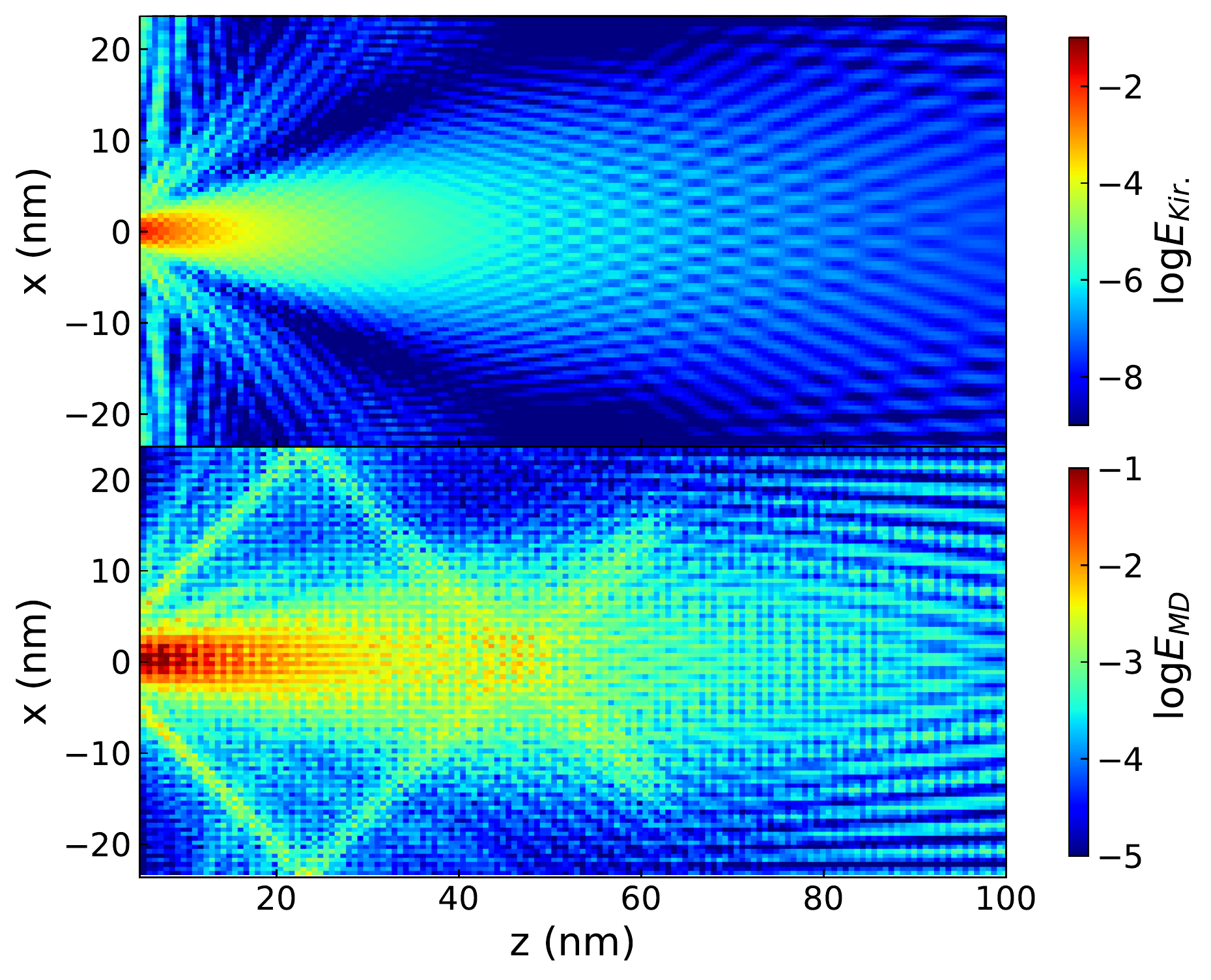}
		
		\caption{ \label{fig:CompLog} Comparison of the non-normalized results obtained with the Kirchhoff integral (top line) and the MD simulation (bottom line). Using a log-scale. } 
	\end{figure}

	This section contains the visual comparison of the patterns obtained with the Kirchhoff integral and the MD simulation in Fig. \ref{fig:CompLog}. It appears clearly that the intensity decreases faster for the Kirchhoff equation compared to MD. This difference may be explained by the focusing of the incoming wave by the channel.
	
	\section{Amorphous Silicon barrier}
	\label{app:a-Si}
\begin{figure}[h] 
		\centering
		\includegraphics[width=.5\linewidth]{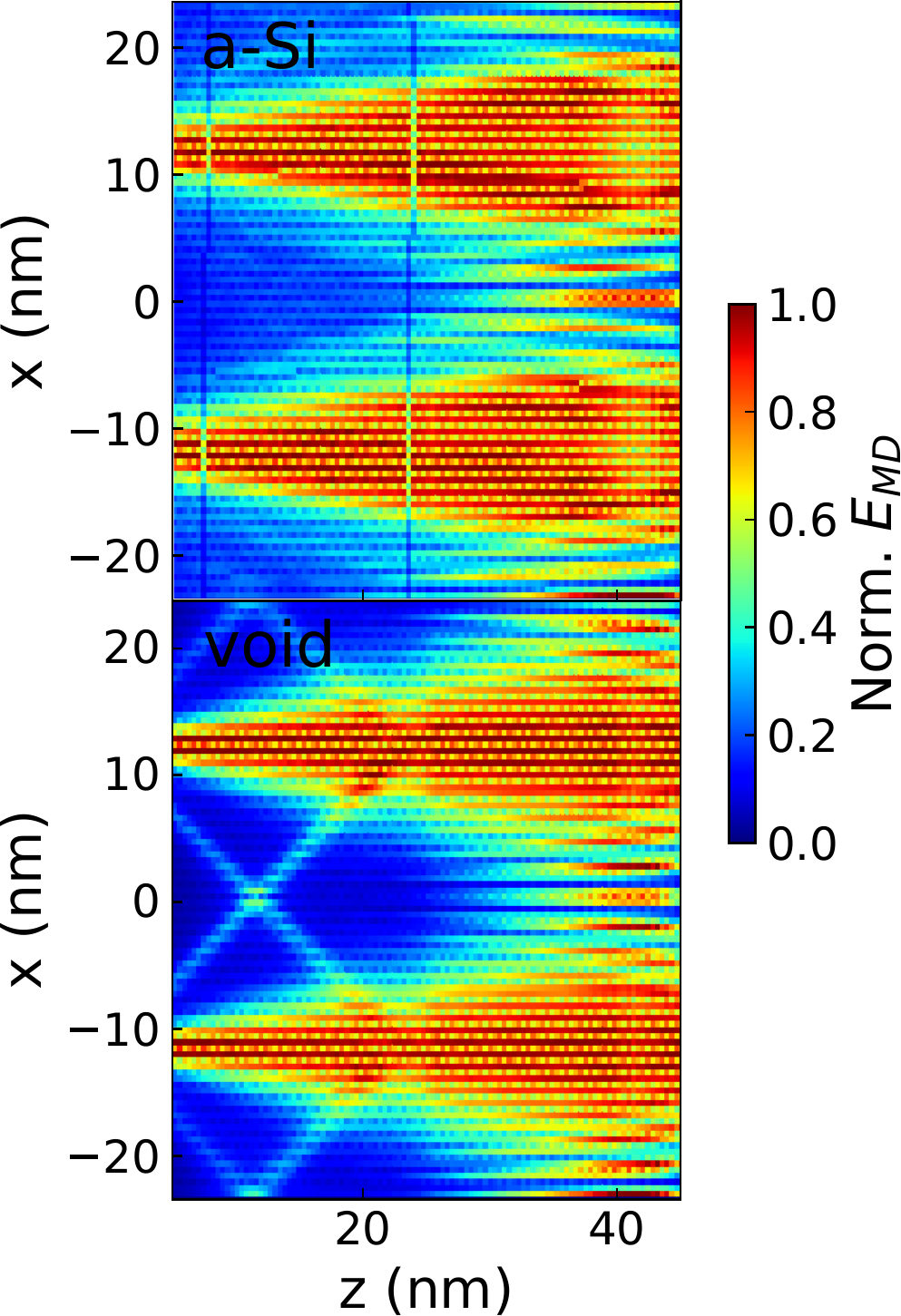}
		
		\caption{ \label{fig:Am} Comparison of the MD simulations, using either a-Si filled screen (top) or a void (top).} 
	\end{figure}

	In this section, the nature of the barrier is studied. For this purpose the results using either a void or an a-Si filled screen are compared. Crystalline window in amorphous barrier can be obtained experimentally \cite{Nakamura2015}.
	The geometry is slightly modified, with two holes being modeled and a distance $d$ of 20.6 nm (see Fig. 2 of the main text) between the aperture opening and the distance between the screen with the apertures and the box limits is also smaller (60 nm). Here, two apertures are simulated and represented. Moreover the excitation is not continuous but a pulse as in a previous work \cite{Desmarchelier2021a} and the surperposition of multiple times steps is used to get the full pattern in figure \ref{fig:Am}.
	Comparing the two panels, it appears that using a-Si results in a broader ray at the exit of the aperture. This probably comes from the a-Si/c-Si interface allowing the transmission of energy \cite{Yang2018}. Overall, there is contrast when using an a-Si barrier. Again, this can be attributed to the partial transmission of energy through the amorphous screen, whereas the void is an insurmountable barrier for the wave.
	Another notable difference is the absence of the secondary fringe at \ang{45} for a-Si. This hints that these fringes are caused by the free surface of the crystalline channel, in which fully specular reflections are possible whereas transmission dominates in the case of a-Si/c-Si interface \cite{Yang2018}. The role of the channel in these side fringes is discussed more in depth in the next section.

	\section{Impact of Crystalline Orientation and Channel Length}
	
		\label{app:or}
		
			\begin{figure}[ht] 
			\centering
			\includegraphics[width=.7\linewidth]{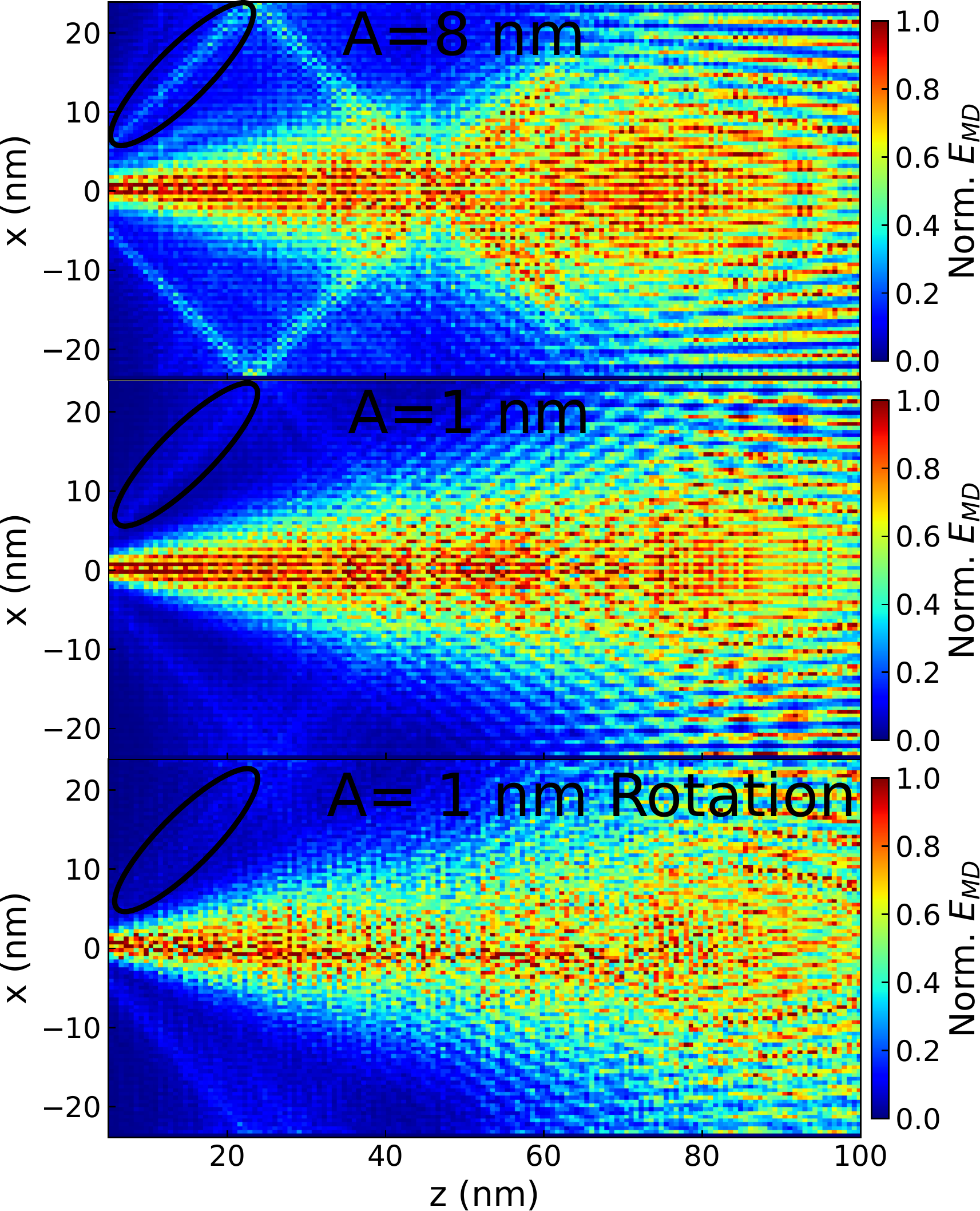}
			
			\caption{ \label{fig:SmallerChannel} Interference pattern as displayed in Fig. 3 of the main text (upper panel), for a channel length $A$ of 1 nm with (lower panel) or without (mid panel) a slight rotation of the crystal.} 
		\end{figure}

	In this section two configurations are introduced for comparison, with a shorter channel: A= 1 nm instead of 8 nm, and for one of them a small rotation of the crystal is introduced. In this last configuration, the $x$ direction is not aligned with the $\langle 100 \rangle$ but with $\langle 50 \ 1 \ 0 \rangle$ to see whether having a high symmetry direction aligned with the propagation direction has an influence on the propagation.
	
	In Fig. \ref{fig:SmallerChannel} these configurations are compared. First, for the shorter channels, the side fringes are less marked (as visible when comparing the regions circled in black). This is consistent with the contribution of the reflections on the channel sides to the appearance of the side fringes.

	If we still focus on this side fringes, it appears that they depend on the crystalline orientation as well. Indeed, when the crystalline orientation is tilted, the upper fringe almost disappears (see the black circled region).
	This indicates that these side fringes depend on high symmetry crystalline orientation. If $x$ is aligned with $\langle 100 \rangle$ they are aligned with $\langle 111 \rangle$.
	Lastly, a small asymmetry in the interference pattern seems to appear upon the rotation, it might be due to the change in orientation. However, one should note that due to the rotation, crystalline defects appear at the periodic boundary conditions in x and this may impact the diffraction patterns.
	\pagebreak

	%
	%
	%

\bibliographystyle{ieeetr}

\bibliography{./BibliClean.bib}{}  

\begin{thebibliography}{10}

\bibitem{Huygens1920}
C.~Huygens, {\em Trait{\'e} de la lumi{\`e}re}.
\newblock Gauthier-Villars, 1920.

\bibitem{Young1802}
T.~Young, ``Ii. the bakerian lecture. on the theory of light and colours,''
  {\em Philosophical Transactions of the Royal Society of London}, vol.~92,
  pp.~12--48, 1802.

\bibitem{Kirchhoff1883}
G.~Kirchhoff, ``Zur theorie der lichtstrahlen,'' {\em Annalen der Physik},
  vol.~254, no.~4, pp.~663--695, 1883.

\bibitem{Deregowski2006}
S.~M. Deregowski and S.~M. Brown, ``A theory of acoustic diffractors applied to
  2‐d models*,'' {\em Geophysical Prospecting}, vol.~31, no.~2, pp.~293--333,
  1983.

\bibitem{Marcella2002}
T.~V. Marcella, ``Quantum interference with slits,'' {\em European Journal of
  Physics}, vol.~23, pp.~615--621, oct 2002.

\bibitem{Toyoda2015}
K.~Toyoda, R.~Hiji, A.~Noguchi, and S.~Urabe, ``Hong--ou--mandel interference
  of two phonons in trapped ions,'' {\em Nature}, vol.~527, no.~7576,
  pp.~74--77, 2015.

\bibitem{Anderson1970}
C.~H. Anderson and E.~S. Sabisky, ``Phonon interference in thin films of liquid
  helium,'' {\em Phys. Rev. Lett.}, vol.~24, pp.~1049--1052, May 1970.

\bibitem{Xie2014}
G.~Xie, Y.~Guo, X.~Wei, K.~Zhang, L.~Sun, J.~Zhong, G.~Zhang, and Y.-W. Zhang,
  ``Phonon mean free path spectrum and thermal conductivity for
  si$_{1-x}$ge$_x$ nanowires,'' {\em Applied Physics Letters}, vol.~104,
  no.~23, p.~233901, 2014.

\bibitem{Hu2020}
S.~Hu, L.~Feng, C.~Shao, I.~A. Strelnikov, Y.~A. Kosevich, and J.~Shiomi,
  ``Two-path phonon interference resonance induces a stop band in a silicon
  crystal matrix with a multilayer array of embedded nanoparticles,'' {\em
  Phys. Rev. B}, vol.~102, p.~024301, Jul 2020.

\bibitem{Simoncelli2019}
M.~Simoncelli, N.~Marzari, and F.~Mauri, ``Unified theory of thermal transport
  in crystals and glasses,'' {\em Nature Physics}, vol.~15, no.~8,
  pp.~809--813, 2019.

\bibitem{Zhang2022}
Z.~Zhang, Y.~Guo, M.~Bescond, J.~Chen, M.~Nomura, and S.~Volz, ``Heat
  conduction theory including phonon coherence,'' {\em Phys. Rev. Lett.},
  vol.~128, p.~015901, Jan 2022.

\bibitem{Hanus2018}
R.~Hanus, A.~Garg, and G.~J. Snyder, ``Phonon diffraction and dimensionality
  crossover in phonon-interface scattering,'' {\em Communications Physics},
  vol.~1, no.~1, pp.~1--11, 2018.

\bibitem{Vu1995}
P.~D. Vu, J.~R. Olson, and R.~O. Pohl, ``Phonon diffraction gratings,'' {\em
  Annalen der Physik}, vol.~507, no.~1, pp.~9--25, 1995.

\bibitem{Dieleman1999}
D.~J. Dieleman, A.~F. Koenderink, A.~F.~M. Arts, and H.~W. de~Wijn,
  ``Diffraction of coherent phonons emitted by a grating,'' {\em Phys. Rev. B},
  vol.~60, pp.~14719--14723, Dec 1999.

\bibitem{Zaitlin1975}
M.~P. Zaitlin and A.~C. Anderson, ``Phonon thermal transport in noncrystalline
  materials,'' {\em Phys. Rev. B}, vol.~12, pp.~4475--4486, Nov 1975.

\bibitem{Desmarchelier2021}
P.~Desmarchelier, A.~Tanguy, and K.~Termentzidis, ``Thermal rectification in
  asymmetric two-phase nanowires,'' {\em Phys. Rev. B}, vol.~103, p.~014202,
  Jan 2021.

\bibitem{Vink2001}
R.~Vink, G.~Barkema, W.~{van der Weg}, and N.~Mousseau, ``Fitting the
  stillinger–weber potential to amorphous silicon,'' {\em Journal of
  Non-Crystalline Solids}, vol.~282, no.~2, pp.~248--255, 2001.

\bibitem{Plimpton1997}
S.~Plimpton, ``Fast parallel algorithms for short-range molecular dynamics,''
  {\em Journal of Computational Physics}, vol.~117, no.~1, pp.~1--19, 1995.

\bibitem{Peatross2015}
J.~Peatross and M.~Ware, ``Physics of light and optics,'' p.~JWA64, available
  at optics.byu.edu, 2015.

\bibitem{SciPy}
P.~Virtanen, R.~Gommers, T.~E. Oliphant, M.~Haberland, T.~Reddy, D.~Cournapeau,
  E.~Burovski, P.~Peterson, W.~Weckesser, J.~Bright, S.~J. {van der Walt},
  M.~Brett, J.~Wilson, K.~J. Millman, N.~Mayorov, A.~R.~J. Nelson, E.~Jones,
  R.~Kern, E.~Larson, C.~J. Carey, {\.I}.~Polat, Y.~Feng, E.~W. Moore,
  J.~{VanderPlas}, D.~Laxalde, J.~Perktold, R.~Cimrman, I.~Henriksen, E.~A.
  Quintero, C.~R. Harris, A.~M. Archibald, A.~H. Ribeiro, F.~Pedregosa, P.~{van
  Mulbregt}, and {SciPy 1.0 Contributors}, ``{{SciPy} 1.0: Fundamental
  Algorithms for Scientific Computing in Python},'' {\em Nature Methods},
  vol.~17, pp.~261--272, 2020.

\bibitem{Hopcroft2010}
M.~A. Hopcroft, W.~D. Nix, and T.~W. Kenny, ``What is the young's modulus of
  silicon?,'' {\em Journal of Microelectromechanical Systems}, vol.~19, no.~2,
  pp.~229--238, 2010.

\bibitem{Thomsen1986}
L.~Thomsen, ``Weak elastic anisotropy,'' {\em Geophysics}, vol.~51, no.~10,
  pp.~1954--1966, 1986.

\bibitem{Wolfe1998}
J.~Wolfe, ``Imaging phonons: acoustic wave propagation in solids. cambridge
  university press,'' 1998.

\bibitem{Anufriev2020}
R.~Anufriev and M.~Nomura, ``Ray phononics: Thermal guides, emitters, filters,
  and shields powered by ballistic phonon transport,'' {\em Materials Today
  Physics}, vol.~15, p.~100272, 2020.

\bibitem{Nakamura2015}
Y.~Nakamura, M.~Isogawa, T.~Ueda, S.~Yamasaka, H.~Matsui, J.~Kikkawa,
  S.~Ikeuchi, T.~Oyake, T.~Hori, J.~Shiomi, and A.~Sakai, ``Anomalous reduction
  of thermal conductivity in coherent nanocrystal architecture for silicon
  thermoelectric material,'' {\em Nano Energy}, vol.~12, pp.~845--851, 2015.

\bibitem{Desmarchelier2021a}
P.~Desmarchelier, A.~Carré, K.~Termentzidis, and A.~Tanguy, ``Ballistic heat
  transport in nanocomposite: The role of the shape and interconnection of
  nanoinclusions,'' {\em Nanomaterials}, vol.~11, no.~8, 2021.

\bibitem{Yang2018}
L.~Yang, B.~Latour, and A.~J. Minnich, ``Phonon transmission at
  crystalline-amorphous interfaces studied using mode-resolved atomistic
  green's functions,'' {\em Phys. Rev. B}, vol.~97, p.~205306, May 2018.

\end{thebibliography}






\end{document}